\documentclass[preprint,12pt,authoryear]{elsarticle}

\usepackage{amssymb}
\usepackage{url}
\usepackage{amsmath} 
\usepackage{adjustbox}
\usepackage{booktabs} 
\usepackage{enumitem}

\setlist[itemize]{noitemsep, topsep=0pt} % Reduces space between items in the itemize list

\usepackage{xcolor} % For color definitions
\usepackage{hyperref} % For handling hyperlinks and citations

\hypersetup{
    colorlinks=true, % Enable colored links
    citecolor=red,  % Set the citation color to red
    linkcolor=blue, % Optional: set the color for internal links
    urlcolor=blue   % Optional: set the color for URLs
}

\journal{Nuclear Physics B}

\begin{document}

\begin{frontmatter}

%% Title, authors and addresses

%% use the tnoteref command within \title for footnotes;
%% use the tnotetext command for theassociated footnote;
%% use the fnref command within \author or \affiliation for footnotes;
%% use the fntext command for theassociated footnote;
%% use the corref command within \author for corresponding author footnotes;
%% use the cortext command for theassociated footnote;
%% use the ead command for the email address,
%% and the form \ead[url] for the home page:
%% \title{Title\tnoteref{label1}}
%% \tnotetext[label1]{}
%% \author{Name\corref{cor1}\fnref{label2}}
%% \ead{email address}
%% \ead[url]{home page}
%% \fntext[label2]{}
%% \cortext[cor1]{}
%% \affiliation{organization={},
%%            addressline={}, 
%%            city={},
%%            postcode={}, 
%%            state={},
%%            country={}}
%% \fntext[label3]{}

\title{Mortality simulations for insured and general populations}

%% use optional labels to link authors explicitly to addresses:
%% \author[label1,label2]{}
%% \affiliation[label1]{organization={},
%%             addressline={},
%%             city={},
%%             postcode={},
%%             state={},
%%             country={}}
%%
%% \affiliation[label2]{organization={},
%%             addressline={},
%%             city={},
%%             postcode={},
%%             state={},
%%             country={}}

\author[label1]{Asmik Nalmpatian}

\affiliation[label1]{organization={Department of Statistics},%Department and Organization
            addressline={LMU Munich},  
            country={Germany}}

\author[label1]{Christian Heumann}

\begin{abstract}
%% Text of abstract
This study presents a framework for high-resolution mortality simulations tailored to insured and general populations. Due to the scarcity of detailed demographic-specific mortality data, we leverage Iterative Proportional Fitting (IPF) and Monte Carlo simulations to generate refined mortality tables that incorporate age, gender, smoker status, and regional distributions. This methodology enhances public health planning and actuarial analysis by providing enriched datasets for improved life expectancy projections and insurance product development.

\end{abstract}
 

\end{frontmatter}

\section{Statement of need}

Detailed and disaggregated mortality simulations are critical for understanding variations in mortality risk across different demographic groups. However, acquiring high-quality, granular mortality datasets is challenging due to privacy restrictions, proprietary control over insurance data, and legal barriers to data sharing. This lack of detailed data affects public health policy, risk assessment, and insurance calculations.

Current efforts, while valuable, often suffer from limited scope, resolution, or are confined to specific demographics. For instance, methodologies for estimating mortality rates from narrow age windows \citep{Goldstein2023}, small-area mortality estimation \citep{Denecke2023}, and COVID-related predictions \citep{Duchemin2022} demonstrate the utility of such approaches but also underscore the inadequacy of existing data for high-resolution research. Further studies have shown the potential of granularity for improving mortality modeling but also highlight the challenges associated with data standardization and accessibility \citep{rki2014mortality, plosone2011mortality, gerontologist1996mortality}. 
 
For insured populations, precise mortality estimates are essential for setting fair premiums, evaluating longevity risk, and designing insurance products that accurately reflect demographic differences. In the absence of comprehensive datasets, actuaries and researchers must rely on aggregated data, leading to potential biases in mortality estimates.

This study introduces a simulation-based framework that overcomes these limitations by generating synthetic but statistically accurate mortality datasets. By enriching mortality tables with demographic covariates, we enable more precise analysis of mortality trends, supporting both public health initiatives and actuarial applications.

\section{Notation}
\label{sec:notation}

In this section, we provide a summary of the notation and symbols used throughout the paper for clarity and ease of reference. Our analyses are based on multi-dimensional demographic cells (e.g., combinations of age, gender, smoker status, region, etc.), which are often indexed using multiple subscripts. To simplify later model specification, we introduce a unified indexing scheme that maps each multi-dimensional demographic subgroup to a single index.

\begin{itemize}
    \item \( x_{ijk} \): Count (e.g., population or deaths) in the demographic subgroup defined by the combination of dimensions \( i \), \( j \), and \( k \). For example, \( i \) might index age groups, \( j \) gender, and \( k \) smoker status.
    
    \item \( x_{ij\cdot} \): Marginal count obtained by summing over the third dimension (e.g., smoker status), i.e., \( x_{ij\cdot} = \sum_k x_{ijk} \).
    
    \item \( \pi_{ijk} \): Joint probability (e.g., population share) associated with subgroup \( (i, j, k) \).
\end{itemize}

To facilitate model estimation, we collapse and re-index the multidimensional demographic structure into a single flat index \( i \), where each value of \( i \) corresponds to a unique combination of categorical levels across all dimensions (e.g., a 40-year-old female smoker in Bavaria). This one-dimensional indexing refers to demographic subgroups — not individual persons — and simplifies notation in subsequent modeling steps such as regression:

\begin{itemize}
    \item \( D_i \): Observed number of deaths in the insured population for demographic subgroup \( i \).

    \item \( D^P_i \): Observed number of deaths in the general population for demographic subgroup \( i \).

    \item \( E_i \): Exposure (e.g., population size or person-years) for subgroup \( i \).

    \item \( \mu_i \): Mortality rate for subgroup \( i \) in the insured population, to be estimated.

    \item \( \hat{\mu}_i \): Estimated mortality rate for subgroup \( i \) in the insured population.

    \item \( f_1(\text{age}_i) \): Smooth function capturing the non-linear effect of age on mortality.

    \item \( f_2(D^P_i) \): Smooth function capturing the relationship between deaths in the general population and mortality in the insured population.

    \item \( \text{gender}_i \times \text{smoker}_i \): Interaction term indicating combined effect of gender and smoking status in the model.
\end{itemize}

Throughout the paper, we use the term *demographic subgroup* to refer to a unique combination of variables such as age, gender, region, and smoker status. When referring to the index \( i \), we mean a specific demographic subgroup (not an individual), and in the context of modeling, we treat each subgroup as one observation unit.

\section{Methodology}

To address the challenge of generating high-resolution mortality data, our methodological framework proceeds in three key stages. It combines demographic inference, synthetic data generation, and advanced statistical modeling to create reliable and granular mortality estimates for both insured and general populations:

\begin{enumerate}
    \item We start by estimating mortality rates using available marginal distributions of demographic variables such as age, gender, and smoker status. Due to limitations in fully observed data, we incorporate known constraints via marginals to approximate mortality across subgroups. 

    \item Using Iterative Proportional Fitting (IPF), we derive joint distributions over the population structure and associated mortality patterns that are consistent with the known marginals. These joint distributions serve as the basis for generating new data via Monte Carlo simulation, where death counts are sampled from Poisson distributions according to the inferred demographic composition.

    \item The simulated datasets are then used to estimate mortality rates with greater granularity. Specifically, we apply Generalized Additive Models (GAMs) with Poisson assumptions and demographic covariates to account for non-linear effects and interactions, enabling flexible and robust predictions even in sparse data settings. This modeling step enables us to infer insured population mortality rates from general population data, particularly in countries where insured-specific data is limited or unavailable.
\end{enumerate}

\subsection{Iterative Proportional Fitting}

IPF is a widely used deterministic method for adjusting contingency tables to match known marginal totals and has been a cornerstone in statistical analysis since its introduction \citep{deming1940least}. It iteratively refines initial estimates to ensure consistency across multiple demographic dimensions while preserving the structure of the observed data. Renowned for its efficiency and robustness, IPF calculates non-integer weights that reflect how representative each individual is within each zone, effectively reweighting the data to align with known marginal totals. This method is particularly advantageous in scenarios requiring the estimation of internal cells of a matrix based on these marginals, as it maximizes entropy by exploring the number of configurations that could yield the same marginal counts \citep{cleave1995entropy}.

The IPF process involves iteratively adjusting an input matrix to ensure that its internal cells align with given marginal totals, which typically represent known values across an entire population. For example, in voter migration analysis, the input matrix might represent voter preferences across different election years, with known marginal totals indicating actual vote distributions. Each iteration of IPF refines the matrix by alternately adjusting row and column totals to match the respective marginal distributions, using Maximum Likelihood estimation to update internal cell values. However, convergence is not always guaranteed, particularly when zero entries are present, necessitating practical constraints such as iteration limits or tolerance thresholds for deviations \citep{pukelsheim2014biproportional}.

In our context, IPF is employed to calculate multi-dimensional distributions essential for population simulations. Given that mortality data comprises populations and deaths within each subgroup, our objective is to determine the joint distribution for each additional variable. For instance, knowing the age and state population distributions, we aim to compute the joint distribution across age and state categories. Consider a multiway table in \(N\) dimensions, each representing a sociodemographic variable. For illustrative purposes, assume \(N = 3\). The multiway table \(\pi_{ijk}\) contains unknown components, subject to constraints defined by marginal distributions \(\{x_{ij\cdot}, x_{i\cdot k}, x_{\cdot jk}\}\). The constraints ensure that the sum of observations in each category matches the known marginals and the total number of observations, \(n\).
The IPF process begins with an initial estimate \(\pi_{ijk}^{(0)}\) and proceeds through iterations to adjust the table according to the given marginals. The algorithm can be extended to higher dimensions, facilitating the synthesis of population data at varying resolutions. For instance, when considering three demographic variables, one iteration of the IPF process can be represented as follows:

\begin{equation}
\pi_{ijk}^{(1)} = \frac{1}{n} \frac{x_{ij\cdot} \pi_{ijk}^{(0)}}{\pi_{ij\cdot}^{(0)}}
\end{equation}

\begin{equation}
\pi_{ijk}^{(2)} = \frac{1}{n} \frac{x_{i\cdot k} \pi_{ijk}^{(1)}}{\pi_{i \cdot k}^{(1)}}
\end{equation}

\begin{equation}
\pi_{ijk}^{(3)} = \frac{1}{n} \frac{x_{\cdot jk} \pi_{ijk}^{(2)}}{\pi_{\cdot jk}^{(2)}}
\end{equation}

Each equation represents an update step where the estimated cell probability \(\pi_{ijk}\) is iteratively adjusted to match the given marginals. Specifically, equation (1) adjusts the initial estimate \(\pi_{ijk}^{(0)}\) to align with the marginal totals \(x_{ij\cdot}\), ensuring consistency along the first dimension. Equation (2) further refines \(\pi_{ijk}\) using the marginal totals \(x_{i\cdot k}\) from the second dimension. Equation (3) completes the iteration by incorporating \(x_{\cdot jk}\), ensuring alignment with the third dimension.

This iterative process continues until convergence, ensuring that the synthesized dataset accurately represents the given marginal distributions across all dimensions \citep{agresti2012categorical}.

Incorporating additional variables, such as smoker status, into mortality risk assessments requires accounting for distinct mortality risks while keeping all other characteristics constant. By applying known hazard ratios for different categories, we can refine mortality tables to reflect these differences accurately. Specifically, we first estimate total deaths using age-gender-specific mortality rates for a hypothetical population of 100.000. Then, using the given hazard ratios, we allocate these deaths proportionally across smoker and non-smoker groups of the same total size. This approach ensures that the original age-gender mortality risks are preserved within each subgroup while maintaining the intended hazard ratio structure.

We implemented our methodology using the \texttt{mipfp} R package. For multi-dimensional interactions (e.g., age-gender, gender-smoker), there are two possible approaches:  

\begin{enumerate}
    \item \textbf{Separate IPF runs:} One option is to run IPF separately for different subgroups (e.g., separately for males and females) while ensuring that each subgroup aligns with the corresponding one-dimensional marginal distributions (e.g., for age, state, and smoker status).  
    \item \textbf{Incorporating cross-tabulated constraints:} Alternatively, the \texttt{mipfp} package allows for directly incorporating interactions by using cross-tabulated marginal distributions (e.g., age-gender bivariate marginals). This approach provides a more compact implementation, reducing the degrees of freedom for the algorithm and enabling faster convergence without compromising accuracy.  
\end{enumerate}

The advantage of including cross-tabulated constraints is that it ensures dependencies between variables are explicitly modeled, which becomes increasingly relevant as the number of interaction dimensions grows. This results in a more efficient and scalable implementation, particularly when dealing with complex dependencies among demographic variables.

In summary, IPF serves as a foundational method for population and death synthesis, enabling the creation of detailed and accurate demographic distributions necessary for high-resolution mortality data simulations.

\subsection{Monte Carlo Simulation}

When analytical solutions are unavailable, Monte Carlo simulations provide a solid alternative by approximating these expectations through the simulation of random processes. Using predefined probability distributions, we generate synthetic mortality scenarios that allow for variability assessment. By averaging the simulated values, we obtain estimates that often closely approximate the true expectations. This approach leverages the principle that sample averages are frequently reliable estimators of their corresponding population expectations \citep{robert2004monte}:

\[
\bar{\theta}_n = \frac{1}{n} \sum_{i=1}^{n} X_i \rightarrow \theta = \mathbb{E}[X]
\]

This convergence is underpinned by the assumption that the data are independent and identically distributed (iid) from a distribution with finite variance. The Central Limit Theorem (CLT) provides the convergence in distribution of the sample mean to a normal distribution:

\[
\sqrt{n}(\bar{\theta}_n - \theta') \xrightarrow{d} \mathcal{N}(0, \sigma^2)
\]

where \(\sigma^2 = \mathbb{E}[X^2] - (\mathbb{E}[X])^2\) represents the variance of the underlying distribution. This theorem is instrumental in constructing approximate confidence intervals for the Monte Carlo error, providing a measure of the reliability of our estimates.

Thus, Monte Carlo simulations are employed in this study to generate repeated samples from Poisson distributions, which are used to model count data such as yearly deaths given population size as exposure. This probabilistic approach allows us to quantify the variability and uncertainty of mortality projections. For a Poisson distribution, the variance is equal to its expected value, which we utilize to assess the dispersion of our mortality estimates. This framework is essential for ensuring that simulated distributions align with empirical observations. While the mean mortality rate remains unchanged, Monte Carlo provides insights into variance, skewness, and extreme outcomes, helping to better understand the probability of rare but significant deviations (tail risks).

\subsection{Generalized Additive Models}

GAMs offer a flexible approach for estimating mortality rates in insured populations by leveraging population-level mortality data and incorporating demographic variables such as age, gender and smoker status. The model assumes that the observed number of insured deaths (\(D_i\)) follows a Poisson distribution, a common choice for modeling count data in mortality studies.

\vspace{1cm}
The GAM framework is specified with Poisson distributional assumption and log-link. 
The use of Poisson regression ensures non-negativity of predicted counts and facilitates interpretability through the log-link function. Incorporating smooth terms enhances the model's ability to capture these patterns while avoiding overfitting. The Poisson framework and GAM methodology are well-established in demographic and actuarial research. Studies such as \citet{mccullagh1989generalized} and \citet{haberman1996actuarial} highlight the use of generalized linear models, including Poisson regression, for mortality analysis. Additionally, \citet{currie2004smoothing} demonstrate the advantages of smoothing techniques for estimating mortality rates in sparse data settings. The inclusion of the offset term, \(\log(E_i)\), ensures that the model predicts mortality rates rather than raw death counts, enabling meaningful comparisons across demographic groups with varying levels of exposure.

\begin{equation}
D_i \sim \text{Poisson}(\mu_i \cdot E_i),
\end{equation}

Thus, the proposed model for expected insured mortality rates \(\hat{\mu}_i\) is as follows:
\vspace{-0.1cm}
\begin{equation}
\log(\hat{\mu}_i) = f_1(\text{age}_i) + f_2(D^P_i) \cdot \text{gender}_i \times \text{smoker}_i + \log(E_i)
\end{equation}
\vspace{-0.1cm}

To ensure reliable estimates in countries where insured mortality rates are unavailable, we train the model on data from the most similar country where insured rates exist. We assume that the ratio between insured and general population mortality rates remains constant across comparable demographic variables between source and target country. If this assumption is difficult to justify, existing research on country similarity scores, based on insurance and mortality characteristics, can provide guidance \citep{Nalmpatian2024}. These scores help identify the most analogous countries for model training and adjustment, thereby improving the robustness of mortality rate predictions. 

Overall, the proposed model provides a robust framework for predicting insured mortality rates by leveraging population-level data and demographic segmentation. Its foundation in Poisson regression and the incorporation of GAM smooth terms make it particularly well-suited for handling the complexities of mortality data.

\section{Application}

To demonstrate the applicability of our methodology in generating granular mortality data for both insured and general population, we explore three distinct scenarios for Germany, Italy, and Switzerland. Mortality data typically consists of exposure (i.e., population) and death counts, and the IPF method can be applied to both. 

Scenario 1 focuses on enhancing demographic precision while assuming uniform mortality rates across states. Scenarios 2 and 3 incorporate an additional mortality risk factor (smoker status) with distinct hazard rates, while Scenario 3 further extends the methodology to general population data by incorporating an insured population adjustment. 

The application begins by selecting relevant demographic variables and loading distributional assumptions from available general population data (Table~\ref{tab:data_sources}), under the assumption that similar patterns apply to insured populations. If specific insured population data is available, it can be directly incorporated to improve accuracy.

\begin{table}[h!]
\centering 
\caption{Overview of data sources for marginal distributions by country}
\begin{adjustbox}{max width=\textwidth}
\begin{tabular}{lccc}
\hline
 & Germany & Italy & Switzerland \\
\hline
Population and deaths by age and gender & \cite{hmd2023} & \cite{hmd2023} & \cite{hmd2023} \\ 
Population by smoker and gender & \cite{Zeiher2017}  &  \cite{Semyonov2012} & \cite{Gmel2017} \\
Population by state & \cite{DestatisPopulation} & \cite{ISTATPopulation} & \cite{BFSPopulation} \\
Hazard rates smokers vs. non-smokers & -- & \cite{Menotti2014} & \cite{McEvoy2012} \\
Base mortality rates (general population) & \cite{hmd2023} & \cite{hmd2023} & \cite{hmd2023} \\
Base mortality rates (insured population) & \cite{DAV2022} & \cite{ANIA2014} & -- \\
\hline
\end{tabular}
\end{adjustbox}
\label{tab:data_sources}
\end{table}

\subsection{Scenario 1: Enhancing population granularity using base insurance mortality tables}
We begin with a base mortality table segmented by age, gender, and smoker status for the insured population in Germany. The objective is to improve demographic precision by incorporating state-level variations while assuming uniform mortality rates across states. Using marginal demographic distributions (age-gender, smoker-gender, and state) along with age-gender-smoker-specific DAV insurance rates, we disaggregate mortality data to the state level using IPF and generate Monte Carlo simulations. This approach enhances granularity without introducing additional mortality risk differentiation and is extendable to other demographic variables. 
This scenario exemplifies a minimal input data case, demonstrating what can be achieved when only marginal population distributions of an additional variable are available. It highlights the capability of IPF to enhance segmentation by adding one extra demographic variable (state), even in the absence of direct state-specific mortality data. Although we do not possess state-specific mortality rates, death counts still vary across states because the mortality rates are applied to state-segmented population distributions, reflecting differentiated demographic patterns. Simultaneously, Monte Carlo simulations assist in quantifying uncertainty in mortality rates by generating confidence intervals that incorporate population segmentation effects. This is particularly crucial for small states, where mortality estimates can be highly uncertain. The Poisson distribution assumes that the variance equals the mean death count, resulting in different variances for each state.

\subsection{Scenario 2: Accounting for distinct mortality risks in addition to population granularity}
Unlike Scenario 1, this scenario introduces an additional dimension of mortality risk differentiation while refining demographic segmentation. Starting with a base mortality table segmented by age and gender for the insured population in Italy, we extend mortality data to include smoker status and state-level variations. We assume that smokers and non-smokers exhibit distinct hazard ratios, requiring separate mortality rate estimates for each group. This enables a more realistic and differentiated mortality structure while preserving demographic precision. In summary, while Scenario 1 uses IPF to refine population segmentation with fixed mortality rates, in Scenario 2, we extend this by disaggregating death counts while keeping the population constant, thereby refining mortality rates segmentation. Of course, if state-specific mortality data were available, it could be directly incorporated. However, the goal of Scenario 1 is to illustrate how demographic refinements alone (without additional mortality data) already add value.

\subsection{Scenario 3: Extending granular mortality data to the general population}
This scenario builds upon Scenario 2 but begins with a base mortality table for the general population instead of the insured population. The objective is to generate an age-gender-smoker-state mortality table for the entire population, not just insured individuals. Additionally, assuming proportional relationships between insured and general populations in both the target (Switzerland) and source (Germany) countries, we employ a GAM with Poisson regression and an offset to infer mortality estimates from the general to the insured population. This approach demonstrates how base population mortality rates can be adjusted to reflect insured-specific risk characteristics.

\vspace{0.5cm}
Beyond pure simulated mortality data, we provide visual analyses to facilitate comparisons between simulated, population, and insured mortality rates across multiple countries. These visualizations offer an intuitive means of evaluating the plausibility and consistency of the simulated rates. Furthermore, the application includes a comprehensive suite of validation tests to ensure data integrity and accuracy in rate calculations. These tests verify the consistency of demographic proportions and hazard ratios, reinforcing the reliability of the simulated datasets and derived insights.

\section{Results}

This section details the outcomes of our study, focusing on the disaggregation of mortality data using IPF and Monte Carlo simulations across various countries. The results are accessible for review and download via an interactive Shiny app dashboard, which includes a 95\% confidence interval. The app, along with the code and datasets, is freely available on \href{https://github.com/asmiknalmpatian/Simulation-of-segmented-mortality-tables}{GitHub}.

For Germany, we disaggregated the population data using known marginal distributions from open sources, assuming that the insurance population mirrors the general population. The distributions for gender-smoker, state, and age-gender are presented in Tables \ref{tab:smoker_gender}, \ref{tab:state_distribution}, and \ref{tab:age_gender_distribution}, respectively.

\begin{table}[h!]
\footnotesize
\centering
\caption{Smoker-gender population distribution}
\begin{tabular}{lcc}
\hline
\textbf{Smoker} & \multicolumn{2}{c}{\textbf{Gender}} \\
\cline{2-3}
       & Female & Male  \\
\hline
Yes    & 20.8      & 27.0   \\
No     & 79.2       & 73.0   \\
\hline
\end{tabular}
\label{tab:smoker_gender}
\end{table}

\begin{table}[h!]
\footnotesize
\centering
\caption{State population distribution}
\begin{tabular}{lc}
\hline
\textbf{State}                  &  \textbf{Population} \\
\hline
Baden-W\"urttemberg      & 13.4              \\
Bayern                 & 15.9              \\
Berlin                 & 4.47              \\
Brandenburg            & 3.05              \\
Bremen                 & 0.817             \\
Hamburg                & 2.26              \\
Hessen                 & 7.58              \\
Mecklenburg-Vorpommern & 1.92              \\
Niedersachsen          & 9.64              \\
Nordrhein-Westfalen    & 21.5              \\
Rheinland-Pfalz        & 4.93              \\
Saarland               & 1.17              \\
Sachsen                & 4.83              \\
Sachsen-Anhalt         & 2.58              \\
Schleswig-Holstein     & 3.50              \\
Thüringen              & 2.51              \\
\hline
\end{tabular}
\label{tab:state_distribution}
\end{table}

\begin{table}[h!]
\footnotesize
\centering
\caption{Age-gender population distribution}
\begin{tabular}{ccc}
\hline
\textbf{Age} & \multicolumn{2}{c}{\textbf{Gender}} \\
\cline{2-3}
       & Female & Male  \\
\hline
20  & 1.439913  & 1.5818712  \\
21  & 1.507098  & 1.6599224  \\
22  & 1.503754  & 1.6463640  \\
23  & 1.483638  & 1.6237836  \\
24  & 1.515971  & 1.6515359  \\
25  & 1.573950  & 1.7035385  \\
26  & 1.609090  & 1.7303510  \\
27  & 1.671727  & 1.7916157  \\
28  & 1.823225  & 1.9605025  \\
29  & 1.809394  & 1.9270780  \\
30  & 1.846709  & 1.9747674  \\
31  & 1.812673  & 1.9276493  \\
32  & 1.790005  & 1.8826642  \\
33  & 1.739514  & 1.8217951  \\
...  & ...       & ...        \\
\hline
\end{tabular}
\label{tab:age_gender_distribution}
\end{table}

Using these distributions, we applied IPF to obtain the joint age-gender-smoker-state distribution. Table \ref{tab:ipf_result} shows a portion of the resulting distribution.

\begin{table}[h!]
\footnotesize
\centering
\caption{Result after IPF: Age-gender-state-smoker population distribution}
\begin{tabular}{ccccc}
\hline
\textbf{Age} & \textbf{Gender} & \textbf{State} & \textbf{Smoker} & \textbf{Population} \\
\hline
20  & M      & Baden-W\"urttemberg & Yes & 0.02852311 \\
21  & M      & Baden-W\"urttemberg & Yes & 0.02993047 \\
22  & M      & Baden-W\"urttemberg & Yes & 0.02968600 \\
23  & M      & Baden-W\"urttemberg & Yes & 0.02927884 \\
24  & M      & Baden-W\"urttemberg & Yes & 0.02977925 \\
25  & M      & Baden-W\"urttemberg & Yes & 0.03071692 \\
26  & M      & Baden-W\"urttemberg & Yes & 0.03120039 \\
27  & M      & Baden-W\"urttemberg & Yes & 0.03230506 \\
28  & M      & Baden-W\"urttemberg & Yes & 0.03535030 \\
29  & M      & Baden-W\"urttemberg & Yes & 0.03474762 \\
30  & M      & Baden-W\"urttemberg & Yes & 0.03560752 \\
31  & M      & Baden-W\"urttemberg & Yes & 0.03475792 \\
32  & M      & Baden-W\"urttemberg & Yes & 0.03394678 \\
33  & M      & Baden-W\"urttemberg & Yes & 0.03284924 \\
...  &...     & ... & ... & ... \\
 
\hline
\end{tabular}
\label{tab:ipf_result}
\end{table}

\begin{figure}[h!]
    \centering
    \includegraphics[width=\textwidth, trim=0 0 0 100, clip]{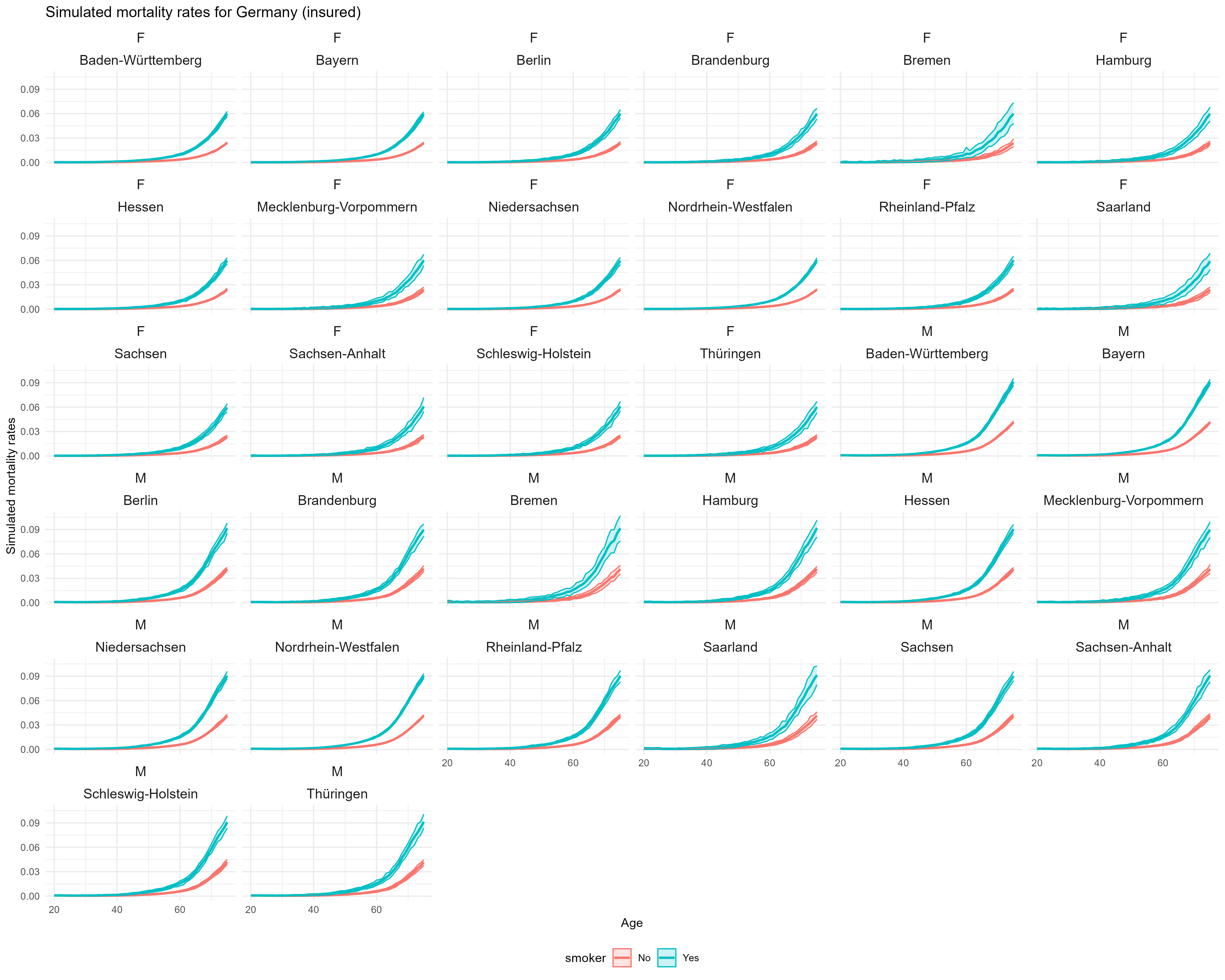} 
    \caption{Simulated mortality rates for Germany.}
    \label{fig:image2}
\end{figure}

Assuming a population size of 1 million, we utilized the derived distribution to estimate expected deaths by applying it to the base mortality table. This involved drawing samples and calculating expected mortality figures, which were then used as inputs for Monte Carlo simulations. Through these simulations, we established 95\% confidence intervals by identifying the 2.5th and 97.5th percentiles of the simulated mortality rates. Figure \ref{fig:image2} provides a detailed visualization of the final mortality rates for Germany, categorized by state, gender, and smoker status. The figure reveals that smaller states exhibit wider confidence intervals, indicating greater variability and uncertainty in mortality estimates due to their smaller population sizes. Smokers demonstrate higher mortality rates compared to non-smokers across all demographic groups, highlighting the essential impact of smoking on mortality. Additionally, males consistently show higher mortality rates than females, underscoring gender as a critical factor in mortality risk assessment. These observed trends are consistent across all states, reflecting our model's ability to account for the distribution of the population across different states. While we assume that the mortality rates themselves are consistent across states, the model adjusts for the proportions of the population within each state. This means that the model effectively captures demographic patterns in mortality by considering how populations are distributed across states. The consistency in trends highlights the ability of our methodology in applying these demographic distributions accurately.

\begin{figure}[h!]
    \centering
    \includegraphics[width=0.8\textwidth, trim=0 0 0 100, clip]{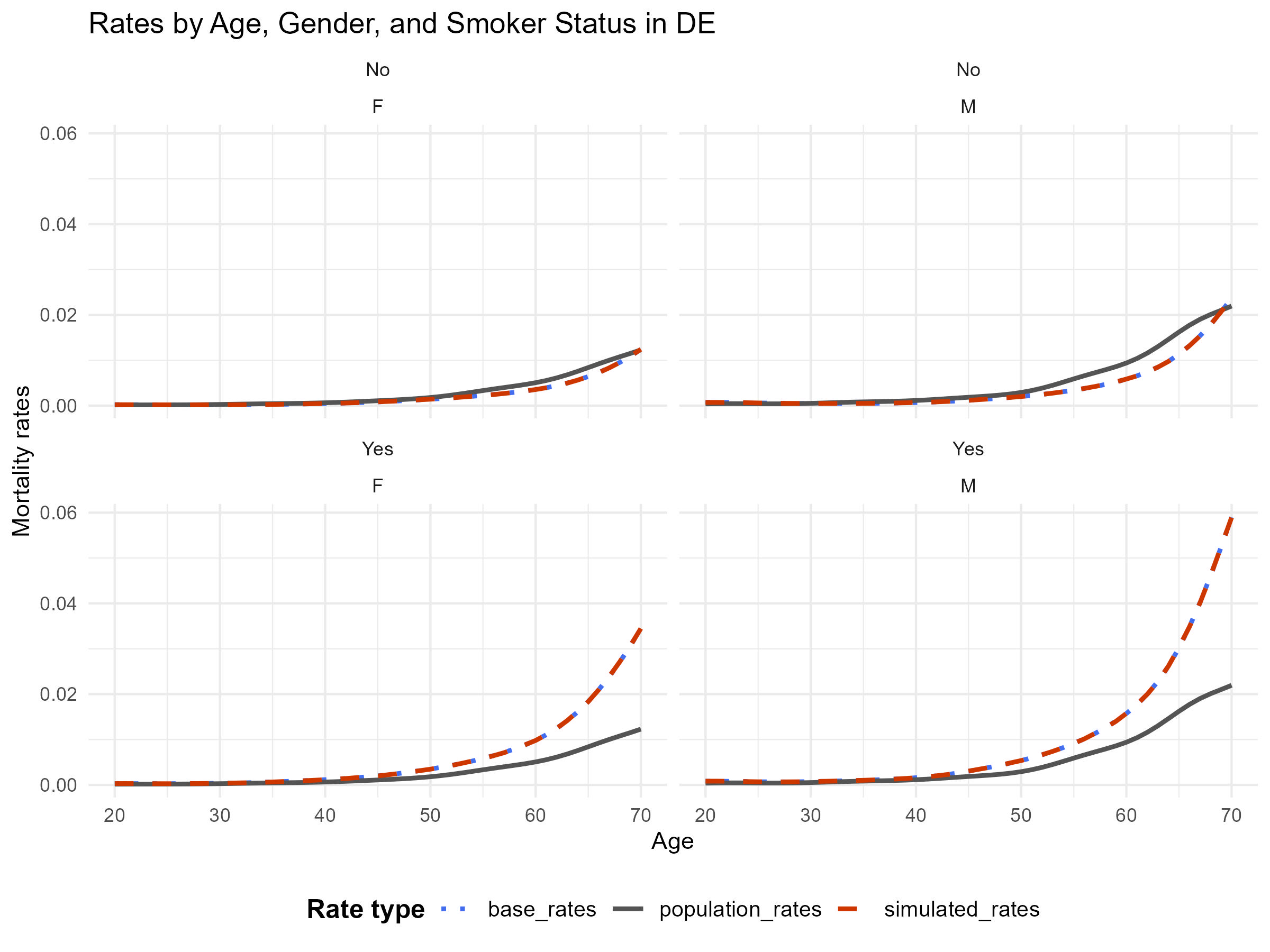} 
    \caption{Aggregated base (insurance), simulated and population mortality rates for Germany.}
    \label{fig:image3}
\end{figure}

Aggregating over all states, Figure \ref{fig:image3} shows that simulated mortality rates align with the base table. For smokers, insurance mortality rates exceed population rates, whereas non-smokers show the opposite trend.

For Italy, since the original base table lacked smoker distinction, we first disaggregated the base mortality table using IPF, starting with age-gender specific mortality data (Table \ref{tab:age_gender_deaths}) from the ANIA insurance population. We applied a hazard ratio of 1.4 to distinguish between smokers and non-smokers, based on the marginal mortality risks (0.014 vs. 0.010). While this ratio was applied uniformly across all subgroups in our primary scenario, the methodology allows for hazard ratios to be specified in a more granular way—varying across age-gender combinations or even higher-dimensional interactions if such detailed information is available.

\begin{table}[h!]
\footnotesize
\centering
\caption{Age-gender mortality rates for insurance population in Italy}
\begin{tabular}{ccc}
\hline
\textbf{Age} & \textbf{Gender} & \textbf{Rates} \\
\hline
20  & M      & 0.000532 \\
21  & M      & 0.000526 \\
22  & M      & 0.000518 \\
23  & M      & 0.000508 \\
24  & M      & 0.000492 \\
25  & M      & 0.000506 \\
26  & M      & 0.000528 \\
27  & M      & 0.000572 \\
28  & M      & 0.000634 \\
29  & M      & 0.000705 \\ 
...  & ...       & ...        \\
\hline
\end{tabular}
\label{tab:age_gender_deaths}
\end{table} 

\begin{figure}[h!]
    \centering
    \includegraphics[width=0.8\textwidth, trim=0 0 0 150, clip]{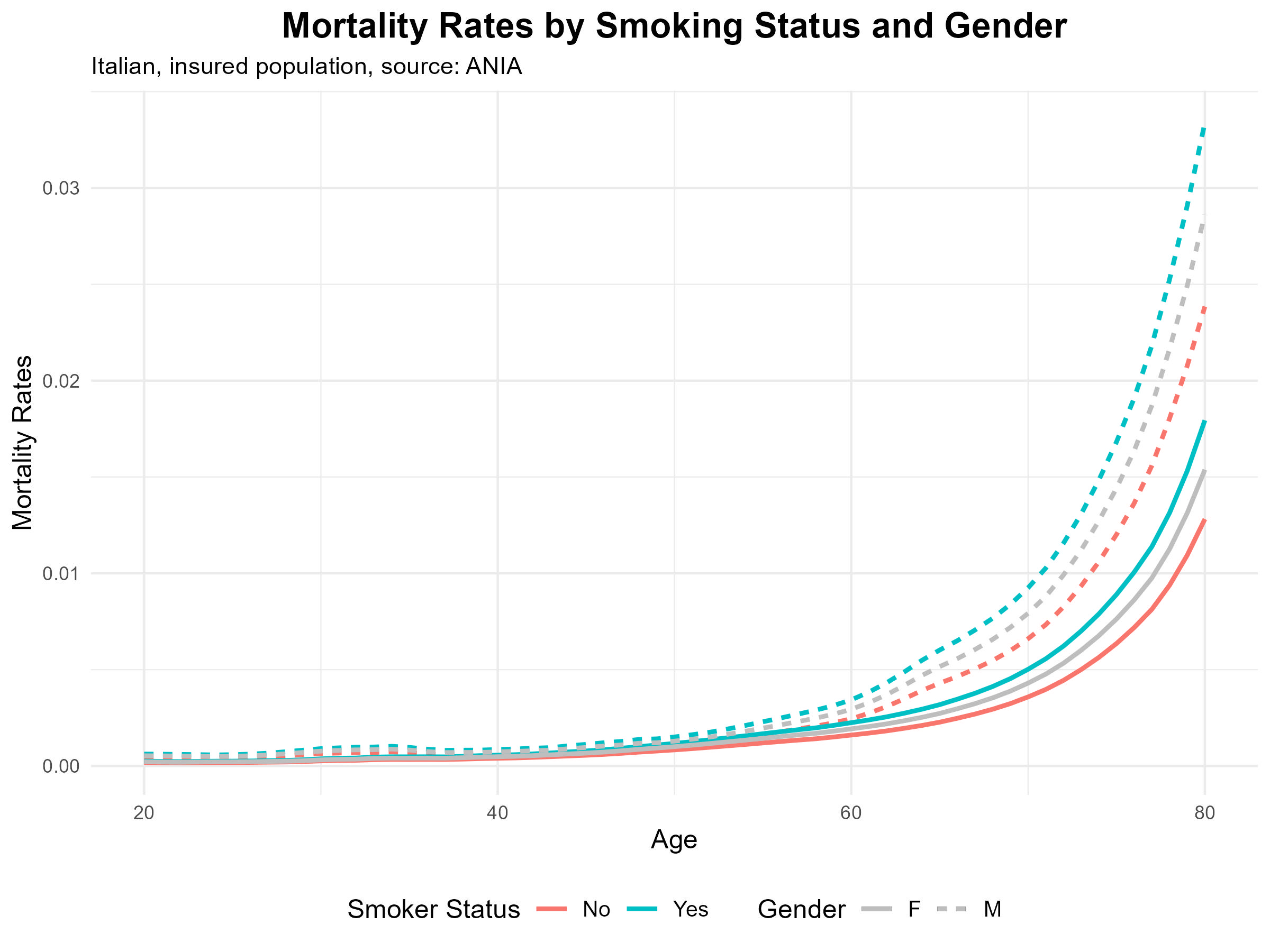} 
    \caption{Disaggregated base mortality table in Italy with IPF.}
    \label{fig:image1}
\end{figure}

The resulting age-gender-smoker mortality rates are shown in Table \ref{tab:result_italy} and Figure \ref{fig:image1}. The curves maintain their shape but shift upwards for smokers and downwards for non-smokers, according to the predefined hazard ratio.

\begin{table}[h!]
\footnotesize
\centering
\caption{Resulting age-gender-smoker mortality rates for insurance population in Italy}
\begin{tabular}{cccc}
\hline
\textbf{Age} & \textbf{Gender} & \textbf{Smoker} & \textbf{Rates} \\
\hline
20  & M      & Yes    & 0.000621 \\
20  & M      & No     & 0.000444 \\
21  & M      & Yes    & 0.000614 \\
21  & M      & No     & 0.000438 \\
22  & M      & Yes    & 0.000604 \\
22  & M      & No     & 0.000431 \\
23  & M      & Yes    & 0.000593 \\
23  & M      & No     & 0.000424 \\
24  & M      & Yes    & 0.000574 \\
24  & M      & No     & 0.000410 \\
\hline
\end{tabular}
\label{tab:result_italy}
\end{table}

\begin{figure}[h!]
    \centering
    \includegraphics[width=0.8\textwidth, trim=0 0 0 100, clip]{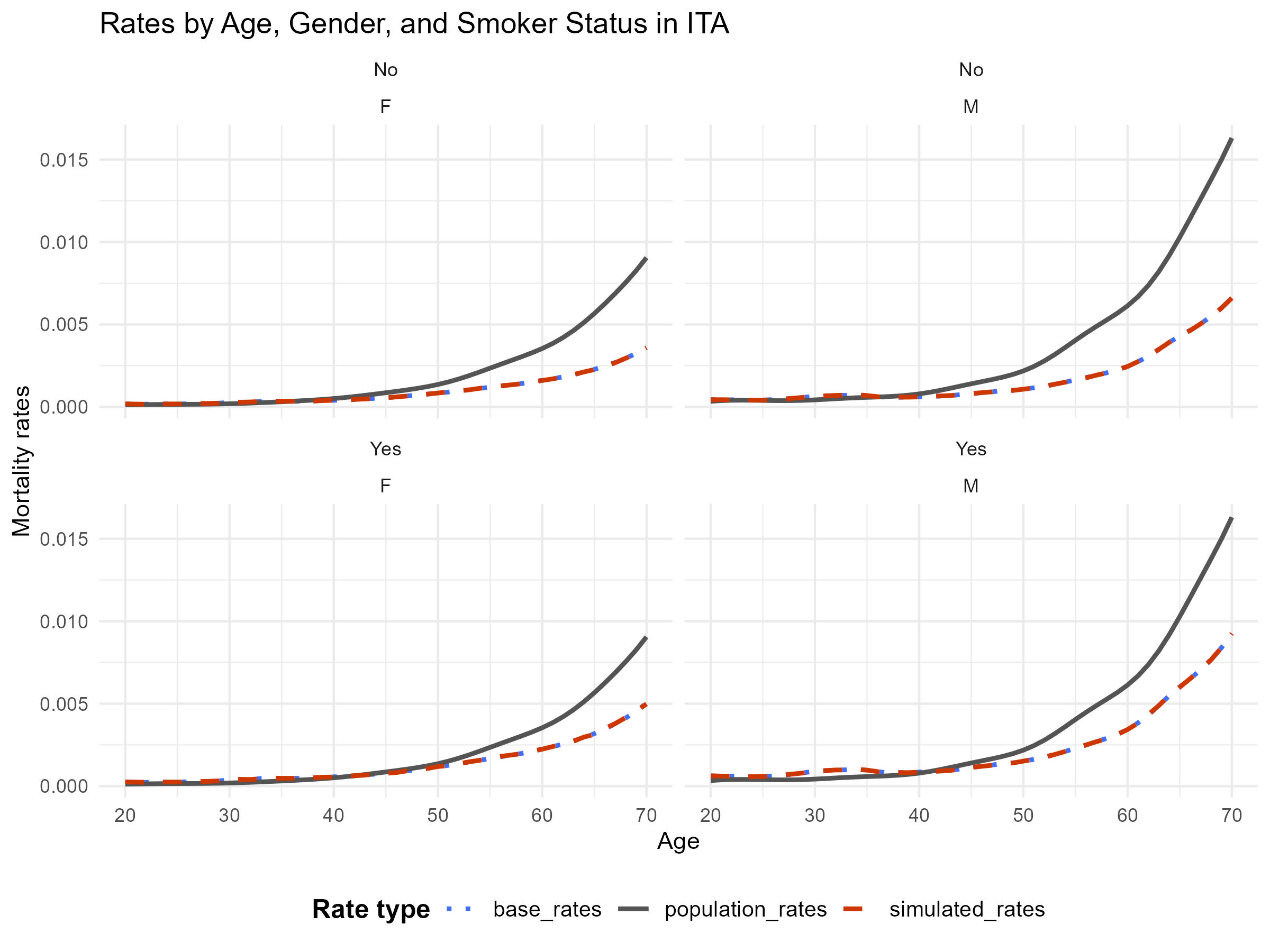} 
    \caption{Aggregated base (insurance), simulated and population mortality rates for Italy}
    \label{fig:image4}
\end{figure}

Figure \ref{fig:image4} demonstrates that, unlike Germany, Italy's population mortality rates for both smokers and non-smokers are generally lower.

\begin{figure}[h!]
    \centering
    \includegraphics[width=0.8\textwidth, trim=0 0 0 100, clip]{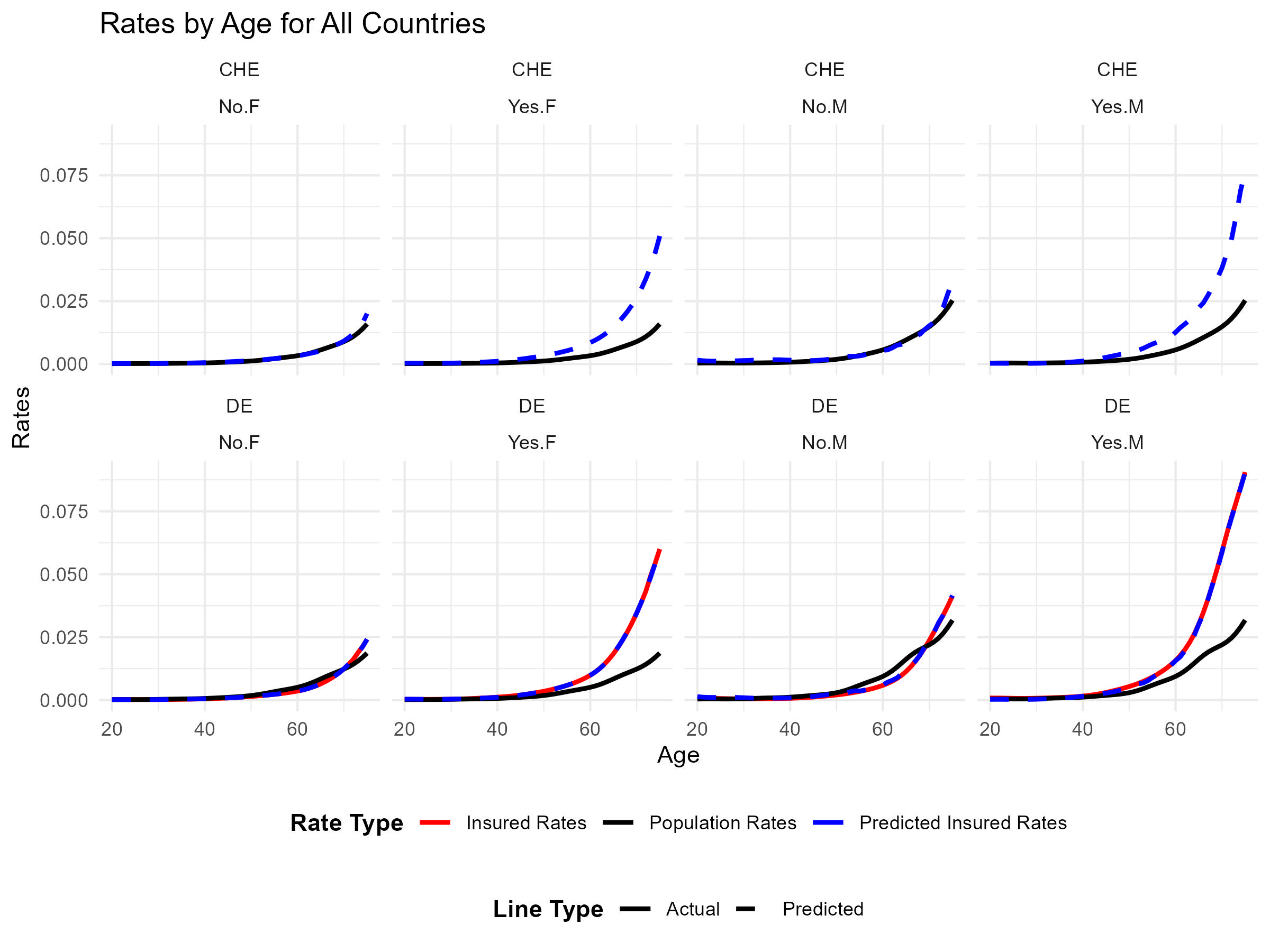} 
    \caption{Inferring insurance mortality for Switzerland based on Germany}
    \label{fig:image5}
\end{figure}

For Switzerland, the base table lacked smoker distinction and was derived from the general population. Disaggregation into smoker and non-smoker categories resulted in distinct mortality curves. Assuming the insured-to-general population ratio mirrors that of Germany, we predicted Swiss population trends, as shown in Figure \ref{fig:image5}. This assumption validates the consistency of our methodology across different national contexts. 

Overall, the results demonstrate the effectiveness of our methodology in disaggregating and analyzing mortality data across different countries, providing valuable insights into population-specific mortality trends.

\section{Limitations}

Our framework lays a strong foundation for mortality simulations in both insured and general populations, howver there are several limitations that present opportunities for future research:

A key limitation in Scenarios 1 and 2 is the potential for selection bias when general population marginals are used in the absence of insured-specific data. Our current approach allows for insured-specific marginals to be inputted when available, which would directly incorporate these differences into the model. However, when such data is unavailable, we use general population marginals as a reasonable approximation. This method acknowledges that some selection effects, such as smoker prevalence, may not be fully captured. An alternative approach could involve adjusting the IPF method to explicitly model selection effects, though this would still require assumptions about the insured distribution if direct data were unavailable. To address the limitations of using general population data, Scenario 3 employs a GAM with Poisson regression. This approach adjusts insured mortality estimates based on observed demographic differences, helping to account for systematic differences between insured and general mortality patterns beyond simple demographic matching. This adjustment highlights the need for more sophisticated modeling techniques when insured-specific marginals are not available.  Future research could integrate additional data sources, like coverage amounts or policy duration, to better model selection effects.

The current model's effectiveness is contingent on the availability and granularity of demographic data. While the methodology allows for extensions to additional demographic variables, the primary challenge remains obtaining sufficiently granular data to support these extensions. For Germany for example, we disaggregate population by state and apply uniform mortality rates, assuming that differences in mortality stem from demographic composition rather than state-specific factors. This simplification overlooks regional disparities in healthcare, environment, or socioeconomic conditions due to the absence of state-level mortality data. Future research could focus on expanding data sources and improving data collection methods.

While Monte Carlo simulations help quantify uncertainty, our approach assumes independent mortality realizations across subgroups. In reality, mortality risks may be correlated across demographic groups, influenced by shared socioeconomic factors. Future work could explore these dependencies to offer more elaborated risk assessments.

Our framework is adaptable to various countries, yet its accuracy hinges on data availability. We have incorporated data from Germany, Italy, and Switzerland, but the quality and granularity of inputs differ across regions. Further validation with additional datasets would be beneficial to assess the approach's generalizability to other markets.

\section{Summary}

In this study, we addressed the challenge of simulating detailed mortality data for both insured and general populations. By integrating multi-dimensional distributional constraints, we employed IPF, enabling the handling of complex demographic interactions and the application of Monte Carlo simulations. Our approach leverages the \texttt{mipfp} R package, facilitating efficient and scalable modeling of population distributions while maintaining accuracy. 

We disaggregated mortality data, including both population and death counts, for Germany, Italy, and Switzerland, taking into account demographic distributions like age, gender, smoker status, and state, along with their interactions. Our findings show that the simulated mortality rates closely match the base tables when aggregated at a higher level. They also provide significant insights into demographic impacts on mortality at a more granular level, generating synthetic insured and general populations while preserving realistic distributional assumptions.

As a prototype, this study presents a robust, privacy-compliant methodology that advances mortality research and actuarial science. Each scenario can be further extended to include more countries, additional variables, or more complex dimensional interactions. The tools and datasets developed are accessible through an open-source interactive dashboard, promoting transparency and further research opportunities. Additionally, the code is available for reproducibility and potential extensions. For an overview of insurance mortality tables from other countries, please refer to the \cite{oecd2023mortality} publication.  

\bibliographystyle{apalike}
\bibliography{references}

\end{document}